\documentclass[11pt]{article} 
\usepackage{hyperref} 
\pdfoutput=1 
\begin{document} 
\title{Sunflower detonation} 
\author{Aslan Kasimov and Svyatoslav Korneev \\ 
\\\vspace{6pt} King Abdullah University of Science and Technology, Saudi Arabia} 
\maketitle 
%% The abstract (in this file, and that submitted as text to arXiv) should 
%include the exact phrase 
%% "fluid dynamics video" or "fluid dynamics videos" 
\begin{abstract} 
In this fluid dynamic video we present simulations of converging two-dimensional detonation in a radially expanding supersonic flow of ideal reactive gas. The detonation is found to be unstable and leads to formation of characteristic cellular patterns. Without any obstacles in the flow,  the detonation keeps expanding radially. To retain the wave within a bounded region, we place a number of rigid obstacles in the flow so that the detonation shock stands some distance toward the center from the obstacles. This leads to generation of reflected shock waves from the obstacles which help the detonation wave to remain at a finite distance from the source. The cellular structure of detonation creates beautiful patterns of shock waves  and contact discontinuities within and after the reaction zone. The patterns  often resemble a sunflower, hence the name of ``sunflower detonation" \cite{kasimov-korneev2012}.
\end{abstract} 
% main text 
\section{Introduction} 

We consider a supersonic two-dimensional flow of an ideal reactive mixture emanating radially out of a source. The flow is described by the reactive Euler equations with a one-step Arrhenius chemistry. It is found that a steady-state circular self-sustained detonation is possible. We compute its structure and investigate its dependence on various parameters of the problem, such as the conditions of the incoming flow and  the mixture thermodynamic and chemical properties. In the simulations shown we chose the parameters to correspond to a mixture of stoichiometric hydrogen-oxygen diluted with helium. Global activation energy, heat release, and the ratio of specific heats are chosen based on those computed by \cite{schultz2000} with detailed chemistry. The rate constant is chosen to impose the unit length of the size of the reaction zone computed by the detailed chemical model in \cite{schultz2000}. Specifically, our simulations correspond to: the helium dilution of $70\%$, the specific-heat ratio of $1.3$, the activation energy of $32.97RT_0$, the heat release of $34.6RT_0$, where $T_0$ is the upstream ambient temperature and $R$ is the universal gas constant. 

The supersonic flow from the source is at a temperature sufficiently low that reactions do not occur, so that the initial expansion is essentially adiabatic. Such flow is known to be accelerating and at some distance, $r_s$, from the source, the flow speed can reach the detonation speed of a given mixture. Therefore, one can expect a standing detonation to exist at about $r_s$. Indeed, the solution of the steady-state reactive Euler equations shows that such detonation exists. It is a self-sustained detonation as there exists a sonic point near the end of the reaction zone. 

To understand the stability of this steady solution, we perform two-dimensional simulations that start with the steady solution as the initial condition. We find that generally, the detonation undergoes the usual cellular instability. However, the detonation structure is seen to enlarge radially so that effectively the wave expands radially out of the computational domain. 
To prevent the unbounded explansion, we placed several rigid obstacles in the flow to be located some short distance behind the steady detonation-shock position. Then several reflected shock waves are formed and the detonation tends to stand, on average, some short distance ahead of the obstacles. This way, the detonation is kept inside the computational domain at least for sufficiently long time.


\begin{thebibliography}{9}

\bibitem{schultz2000}
  E. Schultz, J. Shepherd, "Validation of Detailed Reaction Mechanisms for Detonation Simulation", PhD thesis, California Institute of Technology, 2000.
\bibitem{kasimov-korneev2012}
A. Kasimov, S. Korneev, "On gaseous detonation in a radially expanding flow", G26.00009, Bulletin of the Americal Physical Society, 65th Annual Meeting of the APS Division of Fluid Dynamics, Volume 57, Number 17, San-Diego, CA, 2012.


\end{thebibliography}
\end{document}